\documentclass[12pt]{article}
\usepackage{epsfig}

\def\beq{\begin{equation}}
\def\eeq{\end{equation}}

\def\hybrid{\topmargin 0pt      \oddsidemargin 0pt
        \headheight 0pt \headsep 0pt
        \voffset=1.5cm
       \textwidth 6.5in        
       \textheight 9in         
        \marginparwidth 0.0in
        \parskip 5pt plus 1pt   \jot = 1.5ex}
\catcode`\@=11
\def\marginnote#1{}

\def\numberbysection{\@addtoreset{equation}{section}
        \def\theequation{\thesection.\arabic{equation}}}

\def\underline#1{\relax\ifmmode\@@underline#1\else
        $\@@underline{\hbox{#1}}$\relax\fi}

\def\titlepage{\@restonecolfalse\if@twocolumn\@restonecoltrue\onecolumn
     \else \newpage \fi \thispagestyle{empty}\c@page\z@
        \def\thefootnote{\fnsymbol{footnote}} }

\def\endtitlepage{\if@restonecol\twocolumn \else  \fi
        \def\thefootnote{\arabic{footnote}}
        \setcounter{footnote}{0}}  

\numberbysection
\hybrid

\begin{document}
\title{Bandwidths Statistics from the Eigenvalue Moments for Harper-Hofstadter Problem}
\author{O. Lipan\thanks{%
email : olipan@cco.caltech.edu} \\ Division of Physics, Mathematics, and Astronomy,\\ California Institute of Technology,\\  Pasadena, CA 91125, USA }
 
\maketitle
\begin{abstract}
  I propose a method for studying the product of bandwidths for the Harper-Hofstader model. This method requires knowledge of the moments of the midband energies. I conjectured a general formula for these moments. I computed the asymptotic representation for the product of bandwidths in the limit of a weak magnetic flux using Szeg\" {o}'s theorem for Hankel matrices. I then give a first approximation for the edge of the butterfly spectrum and discuss its connection with P. L\'{e}vy's formula for Brownian motion .
\end{abstract}

\section{Introduction}
 
   A system of electrons on a square lattice in an uniform magnetic field displays an energy spectrum with multifractal properties. Many papers were devoted to this problem. See, for example, \cite {Avron,Harper,Hiramoto,Hofstadter,Jitomirsca,Last, Wannier,Zabrodin,Wilkinson,ThoulessYong1}. A lot of effort has been concentrated to find the Hausdorff dimension of the spectrum at a given magnetic flux. For a rational magnetic flux, i.e. $\Phi =2\pi p/q$, with $p$ and $q$ relatively prime integers, there are $q$ bands in the spectrum. The scaling hypothesis, \cite {Hiramoto}, states that the majority of the bands behaves asymptotically as $q^{-\gamma },$ when $q\to \infty ,$ the rest of the bands being exponetially small. The Hausdorff dimension does not depend on the asymptotic behaviour of the exponentially small bands. The multifractal analysis is concentrated on the bands of the scaling type. The exponentially small bands were studied from a semiclassical aproach in a number of papers. Consult \cite {Helfer} and references therein.

 The aim of this paper is to study the asymptotic properties of the product of bandwidths, so that both scaling and exponentially small bands will manifest themselves in the asymptotic formula. 
 
 First I discuss a method to obtain the band spectrum. Second, I define more precisely the problem I will address.

The Hamiltonian of electrons on a square lattice in a uniform magnetic field, in terms of magnetic translations $T_1$ and $T_2,$ is given by

\beq
\label{Hamil}
H=T_1+T_1^{*}+\lambda (T_2+T_2^{*})\;\;,
\eeq
where $T_1$ and $T_2$ obeys the following commutation relation
\beq
\label{Com}
T_1 T_2=e^{i\Phi}T_2 T_1. 
\eeq 
In (\ref{Com}), $\Phi $ is the magnetic flux. See \cite {Ramal}, \cite {Zabrodin} for details about magnetic translations.
Only the case $\lambda =1$ will be considered hereafter.

For a rational magnetic flux, 
\beq
\label{flux}
\Phi=2\pi\frac{p}{q}\;\;,
\eeq
 the Hamiltonian becomes a $q \times q$ matrix. From now on, $p$ and $q$ will be relatively prime integers.

One representation for the translation operators is 
\beq
\label{Rep1}
T_1=e^{i\theta _1} w_1,\;\;T_2=e^{i\theta _2} w_2\;\;,
\eeq 
where
\beq
\label{w1}
 w_1=\left(\begin{array}{ccccc}
          0      & 1       & 0      & \cdots       & 0 \\
           & \ddots       &  \ddots   & \ddots         &           \\
           \vdots      &   & \ddots    & \ddots       &    0       \\
     
             &  &  & \ddots &1 \\
          1    &           & \cdots      & & 0
         \end{array}\right )\;\;,\eeq
 \beq
\label{w2}
w_2=diag(e^{i\Phi }, e^{i2\Phi },\cdots ,e^{iq\Phi }), 
\eeq
and $\theta _j,\;\;j=1,2 \;$ are real numbers (Bloch parameters).

In this representation,the Hamiltonian is represented by the following matrix:
\beq
\label{MatrHam1}
 H(\theta)=\left(\begin{array}{ccccc}
          a_1      & b_1       & 0      & \cdots       & \bar{b}_q \\
          \bar{b}_1& a_2       &  b_2   &  \cdots       &           \\
          0        & \bar{b}_2 &  a_3   & \ddots       &           \\
     
          \vdots   &  \vdots      & \ddots & \ddots       & b_{q-1}  \\
          b_{q}    &           & 0      & \bar{b}_{q-1}& a_{q}
         \end{array}\right )\;\;,\eeq

where $$a_n(\theta)=2\lambda \cos(\Phi n+\theta _2)\; (n=1,2,\cdots,q)\;\;,$$
      $$b_n(\theta)=e^{i\theta _1}\;\;,$$ and $\theta =(\theta _1,\theta _2).$

 Chamber's relation, \cite {Chambers}, states that the characteristic polynomial of the Hamiltonian matrix decomposes into a $\theta$\ -independent polynomial $P_{p/q}(E)$ of degree $q$ and a function $h(\theta)$.
\beq
\label{Polcar1}
\det(H(\theta)-E)=P_{p/q}(E)+h(\theta)\;\;,
\eeq
\beq
\label{Cham1}
h(\theta)=(-1)^{q-1} 2 (\cos(q \theta _1)+\cos(q \theta _2))\;\;.
\eeq

 When $\theta $ varies, the function $h(\theta )$ varies between $-4$ and $4,$ so, 
for a given rational flux $\Phi=2\pi p/q,$ the bands are obtained by intercepting the graph of the polynomial $P_{p/q}(E)$ with the horizontal lines drawn at $4$ and $-4$. Figure 1 shows the procedure for $p=1,\;\;q=3.$

\begin{figure}[ht]
\centering
\epsfig{figure=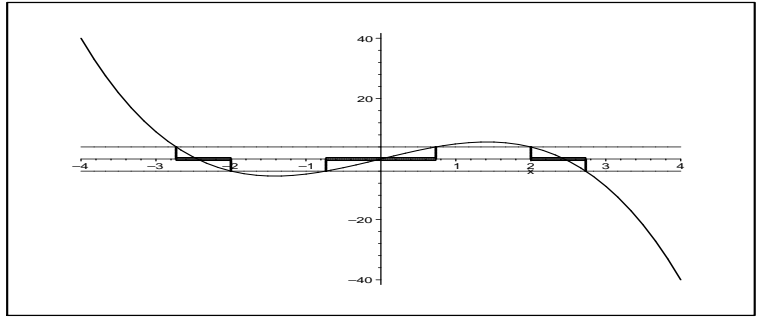,height=10cm, width=4.2cm,angle=270}
\caption{Bands for $P_{1/3}=-x^3+6x$}
\end{figure}

 For the case in the figure $1$ there are three bands, because $q=3$. Counting from left these are $Band_1=[-1-\sqrt 3,-2]$, $\;Band_2=[1-\sqrt 3,-1+\sqrt 3]$, and
$Band_3=[-1+\sqrt 3,1+\sqrt 3 ].$ Denote the widths of these bands $\Delta_1$,
 $\Delta_2$,and  $\Delta_3$ respectivelly. 
In general, for a polynomial $P_{p/q}(E)$, the widths will be named, counting from negative to pozitive energies,  $\Delta_1,\cdots, \Delta_q.$ 
 The roots of the polynomial $P_{p/q}(E)$ are eigenvalues for the  Hamiltonian, for special values of Bloch momenta $\theta=(\pi/q,\pi/q).$
Let us call these roots $E_1,\cdots,E_q,$ counting from the negative to pozitive values of $E$. For the example in figure $1$, $p=1,\;q=3$, these eigenvalues are:
$E_1=-\sqrt 6,E_2=0,E_3=\sqrt 6.$ The energies $E_1,E_2,\cdots, E_q$ are called midband energies. These energies were studied in \cite{ATPaul,Zabrodin} using Bethe-ansatz. 

 The problem can be now stated: find the  asymptotic formula for the products of the bandwidths, i.e. 
\beq
\label{Bandprod}
\prod _{n=1}^{q} \Delta_n .
\eeq
 
 In the last section of the paper, I obtain that, for a week magnetic field $\Phi=2\pi \frac {1}{q}\to 0 ,$ the asymptotic formula can be arranged as a product of terms, each term approaching zero faster than the next neighbour term, (\ref{Prodas}):
\beq
\label{ProdasIntr}
\prod_{n=1}^{q}\Delta _n \sim 2^{-q^2}q^{-q} \exp(-qd_0+d_1+O(1)).
\eeq
Here, $d_0$ and $d_1$ are some constants. We will compare now this result, (\ref{ProdasIntr}),
 to results from
 multifractal analysis of the band spectrum. The multifractal analysis requires the power sums of the bandwidths to be evaluated and the goal is to find the function $F(\beta )$ defined by
\beq
\label{sum}
q^{F(\beta )}=\sum _{n=1}^q \Delta _n^{\beta }\;\;,
\eeq
where $\beta $ is a real parameter.
Taking the derivative with respect to $\beta $ at $\beta =0$ in (\ref{sum}) we get 
\beq
\label{multi}
F^{'}(0)q \ln q=\sum _{n=1}^q \ln \Delta _n\;\;.
\eeq
Compare (\ref{multi}) wih the logarithm of (\ref{ProdasIntr}) and you see that
the term $2^{-q^2}$, which is due to exponentially small bands, does not appear from (\ref{multi}). 
 To incorpoarate this term, we need to know how many bands out of the total number of bands ( which equals $q$) are exponentially small. Introduce a number $f$ between $0$ and $1,$ such that $fq$ is the number of bands that are exponentially small. Than the total number of bands can be decompose as $q=fq+(1-f)q$ and $(1-f)q$ is the number of bands that asymptotically behaves like $q^{-\gamma }.$
 Now the formula (\ref{sum}) can be rewritten as
\beq
\label{sumFG}
fqe^{q G(\beta )}+(1-f) q q^{F(\beta )-1}=\sum _{n=1}^q \Delta _n^{\beta }\;\;.
\eeq
 Here $G(\beta )$ is a function which describes the effect of all  exponetially small bands. 
Than by taking the derivative with respect to $\beta $ at $\beta =0$, but now in (\ref{sumFG}), we get (using $G(0)=0$ and $F(0)=1.$
)
\beq
\label{f}
fG^{'}(0)q^2+(1-f)F^{'}(0)q\ln q=\sum _{n=1}^q \ln \Delta _n\;\;.
\eeq
In this case both leading terms in (\ref{ProdasIntr}) are present in (\ref{f})
 and by comparing the coefficients we can write
\begin{eqnarray}
\label{coef}
(1-f)F^{'}(0)\!\!\!\!&=&\!\!\!\!-1\;\;,\nonumber \\
fG^{'}(0)\!\!\!\!&=&\!\!\!\!-\ln 2\;\;.
\end{eqnarray}
 The above formulae, (\ref{coef}), are valid for the case of a weak magnetic flux $(\Phi \to 0 )$. It will be interesting to find the fraction $f$ as a function of the flux.
 By the scaling hypothesis, the fraction $f$ is close to $0.$ 
I want to emphasise that the formula (\ref{sumFG}) is a hypothesis and work remains to be done to understand the behaviour of $\sum _{n=1}^q \Delta _n^{\beta }.$ 
 See \cite{Hiramoto} for a detailed study of multifractal properties and the bibliography therin.  
   
 Before ending this  section, I write 
the  polynomial $P_{p/q}(E)$ as a characteristic polynomial of a tridiagonal matrix. In this way it will become easier to compute the moments of the roots of the polynomial $P_{p/q}(E)$, which we need to obtain the asymptotic formula (\ref{ProdasIntr}). The moments will be the main object of study in Section $3.$
 From this point to the end of the Introduction, the formulas are taken from \cite{Kreft}. 

 The roots of $P_{p/q}(E)$ are special eigenvalues and correspond for those $\theta$ for which $h(\theta)=0$. 
 Studying the polynomial $P_{p/q}(E)$ becomes more facile 
if we choose another representation for the Hamiltonian matrix :
\begin{eqnarray}
T_1\!\!\!\!&=&\!\!\!\!e^{i\theta _1} w_1\;\;,\\
T_2\!\!\!\!&=&\!\!\!\!e^{i\Phi/2}e^{i\theta _2}e^{i\theta _1} w_2 w_1\;\;.
\end{eqnarray}
 In this case the Hamiltonian matrix acquires the form: 
\beq
\label{MatrHam2}
\tilde{H}(\theta )= \left(\begin{array}{ccccc}
          0     & a_1       & 0      & \cdots       & \bar{a}_q \\
          \bar{a}_1& 0      &  a_2   &  \cdots       &           \\
          0        & \bar{a}_2 &  0   & \ddots       &  0         \\
     
          \vdots   &  \vdots      & \ddots & \ddots       & a_{q-1}  \\
          a_{q}    &           & 0      & \bar{a}_{q-1}& 0
         \end{array}\right )\;\;,
\eeq

with
$$a_r=(1+e^{i\Phi (r+\frac{1}{2})+i\theta _2}) e^{i\theta _1}\;\;.$$

Chamber's relation in the above representation is:
\beq
\label{Polcar2}
\det(\tilde {H}(\theta)-E)=P_{p/q}(E)+\tilde{h}(\theta)\;\;,
\eeq
where
\beq
\label{Cham2}
\tilde{h}(\theta)=(-1)^{q-1} 2 (\cos(q \theta _1)+(-1)^{p}\cos(q (\theta _2+\theta _1)))\;\;.
\eeq
 
 This last representation is convenient because for the special value  $\theta _0=( \theta _{10},\theta _{20})=\\ (0,\pi (1+p/q)),$ simultaneously  
\beq
\label{Zero1}
\tilde{h}(\theta_0)=0
\eeq and
\beq 
\label{Zero2}
a_q(\theta_0)=0\;\;,
\eeq
which makes $\tilde{H}(\theta_0)$ a tridiagonal matrix.
 Out of this tridiagonal matrix, recursive equations can be deduced for generating the polynomial $P_{p/q}(E).$
Let ${\tilde P}_0(E),{\tilde P}_1(E),\cdots,{\tilde P}_q(E)$ be a sequence of polynomials generated by the following relations:
\begin{eqnarray}
 {\tilde P}_0(E)\!\!\!\!&=&\!\!\!\!1\;\;,\label{P_0}\\
 {\tilde P}_1(E)\!\!\!\!&=&\!\!\!\!-E\;\;,\label{P_1}\\
 {\tilde P}_n(E)\!\!\!\!&=&\!\!\!\!-E {\tilde P}_{n-1}(E)-\beta_{n-1} {\tilde P}_{n-2}(E) \;\;\;\;(n=2,3,\cdots,q)\;\;,\label{P-n}
\end{eqnarray}
where
\beq
\label{beta} 
\beta_n=4\sin ^2 \left(\frac {\Phi }{2} n\right)\;\;\;\;(n=1,2,\cdots,q)\;\;,
\eeq 
$$\Phi=2\pi p/q\;\;.$$
 It follows that  ${\tilde P}_q(E)=P_{p/q}(E)$ and that the polynomial $P_{p/q}(E)$ is the characteristic polynomial of the following tridiagonal matrix, which contains only positive numbers:
\beq
\label{M}
M= \left(\begin{array}{ccccc}
          0     & \beta_1       & 0          & \cdots       & 0 \\
          1     & 0             &  \beta_2   &  \cdots       & \\                                                                   0     & 1             &  0          & \ddots       &  0 \\        
     
          \vdots   &  \vdots      & \ddots & \ddots & \beta_{q-1} \\ 
           0   &           & 0      & 1& 0
         \end{array}\right )\;\;,
\eeq
\beq
\label{Polcar3}
P_{p/q}(E)=\det(M-E)\;\;.
\eeq

\section{Bandwidths' Product}

From the paper  \cite{Yoram}, it is known that the bandwidths obey the inequalities (with $e=\exp (1)=2.718\cdots $):
\beq
\label{Yoramineq}
\frac{2(1+\sqrt 5)}{\vert P_{p/q}^{'}(E_n)\vert} < \Delta _n < \frac{8e}{\vert P_{p/q}^{'}(E_n)\vert},\;\;\;n=1,\cdots,q\;\;,
\eeq 
where $P_{p/q}^{'}(E_n)$ is the derivative of $P_{p/q}(E)$ computed at the eingenvalue $E_n,$ which are the roots of $P_{p/q}(E).$
 As a consequence,
\beq
\label{Prodineq}
(2+2\sqrt 5)^{q}\frac{1}{\vert \prod_{n=1}^{q}P_{p/q}^{'}(E_n)\vert} < \prod_{n=1}^{q}\Delta _n < (8e)^{q}\frac{1}{\vert \prod_{n=1}^{q}P_{p/q}^{'}(E_n)\vert}\;\;.
\eeq
The product $\prod_{n=1}^{q}P_{p/q}^{'}(E_n)$ can be reexpressed as:
\beq
\label{Vanderprod}
\prod_{j>i=1}^{q}(E_j-E_i)^2\;\;.
\eeq 
 The formula (\ref{Vanderprod}) can be written as a product of two Vandermonde determinants.\begin{eqnarray}
\prod_{j> i=1}^{q}(E_j-E_i)^2&=&
     \left|\begin{array}{cccc}
          1    & 1      &  \cdots       &  1\\
          E_1     & E_2  &  \cdots  &  E_q   \\                                                                                                E_1^2 & E_2^2   &  \cdots  &   E_q^2 \\        
     
          \vdots   &  \vdots      &  & \vdots  \\ 
           E_1^{q-1}  & E_2^{q-1}  & \cdots      & E_q^{q-1}
         \end{array}\right |
         \left|\begin{array}{ccccc}
          1     & E_1      & E_1^2          & \cdots       & E_1^{q-1} \\
          1     & E_2      & E_2^2   &  \cdots       & E_2^{q-1} \\                                                                   \vdots     & \vdots            &  \vdots         &        &  \vdots \\        
     
          1  &  E_q      & E_q^2& \cdots & E_q^{q-1} 
          
         \end{array}\right |\;\;.
\end{eqnarray}
By this observation, the product $\prod_{n=1}^{q}P_{p/q}^{'}(E_n)$ can be written in terms of the eingenvalue moments, defined as
\beq
\label{Moments}
s_{2k}=\sum_{n=1}^{q}E_n^{2k},\;\;\;k=1,2,3\cdots\;\;.
\eeq

 Putting it all together, it results:
\beq
\label{ProdBand} 
(2+2\sqrt 5)^{q}\frac{1}{\sigma_{q-1}} < \prod_{n=1}^{q}\Delta _n < (8e)^{q}\frac{1}{\sigma_{q-1}}\;\;,
\eeq
where $\sigma_{q-1}$ is the Hankel determinant:
\beq
\label{Hankel}
\sigma_{q-1}=\left|\begin{array}{ccccc}
          q     & s_1       & s_2          & \cdots       & s_{q-1} \\
          s_1     & s_2      & s_3      &  \cdots   &  s_{q}        \\                                                                   s_2     & s_3            &  s_4          & \cdots       &  s_{q+1} \\        
     
          \vdots   &  \vdots      & \vdots &  & \vdots \\ 
           s_{q-1}   & s_q          & s_{q+1}     & \cdots& s_{2q-2}
         \end{array}\right |\;\;.
\eeq

 To get the asymptotic representation of the product of the bandwidths, as $q \to \infty,$
a separate study of the eigenvalue moments, $s_{2k},$ is necessary. 
The next section is dedicated to this study.

\section{Eigenvalues' Moments}
 In the begining of this section, I am  going to show that
\beq
\label{Mometformula}
s_{2k}=q\sum_{j=0}^{[k^2/4]}a_j(k)\cos(2\pi j p/q)\;\; if\;\;\; q>k\;\;,
\eeq
where $a_j$ are integer numbers that do not depend on $p$ and $q.$ Here $[k^2/4]$ is the integer part of $k^2/4.$

 To see this, note that the eigenvalue moments can be computed from the trace of the powers of the matrix $M$, (\ref {M}): 
\beq
\label{Trace}
s_{2k}=Tr(M^{2k})\;\;.
\eeq
 Let us consider two examples, for  $k=4$ and $k=6\;\;:$
\beq
\label{Trace4}
Tr(M^4)=\sum_{m=1}^{q-1}(2\beta _m^2+4\beta _{m-1}\beta _m)\;\;,
\eeq
\beq
\label{Trace6}
Tr(M^6)=\sum_{m=1}^{q-1}(6\beta _m^3+6\beta _{m-1}^2\beta _m+6\beta _{m-1}\beta _m^2+6\beta _{m-2}\beta _{m-1}\beta _m)\;\;,
\eeq
where $\beta _m=0$ if $m <0$. The sum can be extended to $q$, because $\beta _q=0.$ We get then (here $C:=e^{2\pi i p/q}$):
\begin{eqnarray}
\label{M4}
s_4\!\!\!\!&=&\!\!\!\!Tr(M^4)\!=\!\sum_{m=1}^{q}(2(2-C^m-C^{-m})^2+4 (2-C^{m-1}-C^{-m+1})(2-C^m-C^{-         m}))\\\nonumber
       \!\!\!\!&=&\!\!\!\!4q(7+2\cos(2\pi p/q))\;\;.
\end{eqnarray}

In general,
\beq
\label{Mometsbeta}
s_{2k}=\sum_{m=1}^{q}\sum_{l_0+l_1+\cdots+l_j=k}d_{l_0l_1\cdots l_j}\beta _{m-j}^{l_j}\cdots \beta _{m-1}^{l_1}\beta _{m}^{l_0}\;\;,
\eeq
where $j$ takes values between $0$ and $k-1$; compare with (\ref{Trace6}).
The coefficients $d_{l_0l_1\cdots l_j}$ do not depend on $p$ and $q$. They only depend on $k$ and on $(l_0,l_1\cdots l_j).$ Moreover, we can write  
\beq
\label{Egalitate}
\sum_{m=1}^{q}\beta _{m-j}^{l_j}\cdots \beta _{m-1}^{l_1}\beta _{m}^{l_0}=\sum _{n}g_n C^{\xi _n}\;\;,
\eeq
where $g_n$ are integer numbers which do not depend on $p$ and $q$ and the index $n$ runs from $0$ over a finite range. In the above sum, the only nonzero terms will come for those powers $\xi _n$ which do not depend on $m.$ This fact is only a consequence of the structure of $(l_0,l_1,\cdots,l_j)$ and not on $p$ and $q.$ The sum over $m$ will collect $q$ identical terms so in the result the parameter $q$ will be factor out.
Note that for a fixed $k$ the same $(l_0,l_1,\cdots l_j)$-structure is present for all $q>k.$ For $q\leq k$, some $(l_0,l_1\cdots l_j)$ are not present. For example, for $q=k$  the term $\beta _{m-k+1}\cdots \beta _{m-1}\beta _{m}$ is not present, whereas for all $q>k$ is present.
 I end up here the motivation for (\ref{Mometformula}). The maximum value for $j$ in (\ref{Mometformula}) is $[k^2/4].$ This will be explained later, in section $4$, after formula (\ref{area}).
  
 To proceede further, note that 
 in the formula for $s_{2k}$, (\ref{Mometformula}), the parameter $p$ appears only in the ratio $p/q.$ Because of this,
the analysis of the eigenvalue moments can be extended to all real magnetic fluxes. Indeed, for a irational $\omega$, let us consider a sequence of rational approximants $p_n/q_n \to \omega$, as $n\to \infty .$ For each $n$ we get 
\beq
\label{s4}
s_{4,n}=4q_n(7+2\cos (2\pi p_n/q_n))\;\;.
\eeq
 So, in the limit $n\to \infty $, 
\beq
\label{E4}
<E^4(\omega)>:=\lim_{n\to \infty}\frac{\sum_{j=1}^{q_n}E_{j,n}^4}{q_n}=4(7+2\cos(2\pi\omega))\;\;,
\eeq
where $E_{j,n},\;j=1\cdots q_n$ are the special eigenvalues (roots of $P_{p_n/q_n}(E)$) for the flux $\Phi_n=2\pi p_n/q_n.$
 Here are the first five limit eigenvalue averages:

\begin{eqnarray}
\label{eigenmoments}
<E^2(\omega)>\!\!\!\!&=&\!\!\!\! 4\;\;,\\ \nonumber \\
<E^4(\omega)>\!\!\!\!&=&\!\!\!\! 4(7+2\cos(2\pi \omega))\;\;,\\ \nonumber \\
<E^6(\omega)>\!\!\!\!&=&\!\!\!\!4(58+36\cos(2\pi \omega)+6\cos(4\pi \omega))\;\;,\\ \nonumber \\
<E^8(\omega)>\!\!\!\!&=&\!\!\!\!4(539+504\cos(2\pi \omega)+154\cos(4\pi \omega)+24\cos(6\pi \omega)+\\\nonumber
\;\;\;\;\;\;&\;\;\;&\!\!\!\!
4\cos(8\pi \omega))\;\;,\\ \nonumber \\
<E^{10}(\omega)>\!\!\!\!&=&\!\!\!\!4(5486+6580\cos(2\pi \omega)+2770\cos(4\pi \omega)+780\cos(6\pi \omega)+ \\\nonumber
\;\;\;\;\;\;&\;\;\;&\!\!\!\!210\cos(8\pi \omega)+40\cos(10\pi \omega)+10\cos(12\pi \omega))\;\;.
\end{eqnarray}

 Now, a  natural question comes to mind: is there a general formula for $s_{2k}$, or equivalently for $<E^{2k}(\omega)>?$

  At this time, I have a partial answer to this question. From the first eigenvalue moments the following statement can be infered:

 \begin{eqnarray}
\label{Formula}
<E^{2k}(\omega)>=\left (\;^{2k}_{\;k}\right)^2\left (1+\!\sum_{m=1}^{[k^2/4]}k\frac{(2k-2m-1)!!}{(2k-1)!!} \frac {k!}{(k-j_m-2)!} P_m(k)\sin^{2m}(\pi\omega)\right ),
\end{eqnarray}

where for each $m$ the positive integer number $j_m$ is that one for which :
\beq
\label{JM}
2[(j_m+1)^2/4]+2\leq 2m\leq 2[(j_m+2)^2/4]\;\;.
\eeq
 The first pairs $(2m,j_m)$ are
$(2,0),(4,1),(6,2),(8,2),(10,3),(12,3),(14,4),(16,4),(18,4).$\\
 $P_m(k)$ is a polynomial of degree $3 m-j_m-3.$  This is so because the power of
 $k$ in front of $\sin^{2m}(\pi p/q)$ is $2m.$ There is one more detail about $P_m(k)$
 which can be found by inspection of the first few terms. Namely, the coefficients of the largest power in
 $k$, i.e. $\lambda _m $ in :
\beq
\label{PowerP}
P_m(k)=\lambda _m k^{3m-j_m-3}+\cdots \;\;,
\eeq
are generated by
\beq
\label{Generated}
\frac{\sqrt 2 x}{\sinh(\sqrt 2 x)}=\sum_{m=0}^{\infty}\lambda _m x^{2m}\;\;.
\eeq 
 The following is a list of the first three polynomials $P_m.$ The first eight polynomials are listed in the Appendix.

\begin{eqnarray}
P_1(k)\!\!\!\!&=&\!\!\!\! -\frac{1}{3}\;\;,\\ \nonumber \\
P_2(k)\!\!\!\!&=&\!\!\!\!\frac {1}{90}(7k^2-4k-15)\;\;,\\ \nonumber \\
P_3(k)\!\!\!\!&=&\!\!\!\!-\frac {1}{1890} (31k^4-60k^3-101k^2+66k+280)\;\;.
\end{eqnarray}
 From the formula (\ref{Formula}) we get $s_{2k}$ that we need in (\ref{Hankel}).
Namely
\beq
\label{sk}
s_{2k}=q<E^{2k}(p/q)>\;\;\;\; if \;\;\; q>k\;\;.  
\eeq
 In order to compute the asymptotic representation of the Hankel determinant (\ref{Hankel}), a knowledge of $P_m(k)$ will suffice, because the largest 
moment needed in (\ref{Hankel}) is $s _{2q-2}$ which is still given by the formula (\ref{sk}), since $q>(2q-2)/2.$ Because there is no a general formula for the  polynomials $P_m(k),$ the asymptotic representation for the
 bandwidths' product for all magnetic fluxes cannot be computed.

 Still, the case $\Phi=2\pi \frac{1}{q}$ when 
$q\to \infty$ can be analized, which is done in the last section. Before that, the eingevalues' moments can be further studied, due to summation formula (\ref{PowerP}). The anlysis from the next section, besides a value of its own, it is also useful for understanting the asymptotic formula for the bandwidths' product for the case of a week magnetic flux.

\section {The Edge of the Butterfly Spectrum and L\'{e}vy's Formula for Brownian Motion}

In order to obtain a first approximation for the edge of the spectrum, 
we can make use of the general formula which gives the eigenvalue moments (\ref{Formula}).
Let us write the formula for the eigenvalue moments as a power series in $k$ and retain only the largest power in $k$. Call the result the first approximation to the eigenvalues' moments, and use $(1)$ as a superscript  to mark it:

\begin{eqnarray}
\sum_{j=1}^{q_n}E_{j,n}^{2k,(1)}= q_n\left (\;^{2k}_{\;k}\right)^2\sum_{m=0}^{[k^2/4]}\lambda _m \frac {1}{2^{m}}k^{2m} \sin ^{2m}\left(\pi \frac {p_n}{q_n}\right )\;\;.
\end{eqnarray}
 
For $k \to \infty $ we get

\begin{eqnarray}
\label{first}
\sum_{j=1}^{q_n}E_{j,n}^{2k,(1)}=q_n(C_{2k}^{k})^2 \frac {k \sin (\pi \frac {p_n}{q_n})}{\sinh \left(k \sin (\pi \frac {p_n}{q_n})\right )}\;\;,
\end{eqnarray}
and 

\begin{eqnarray}
E_{max}^{(1)}(\omega )=\lim _{p_n/q_n\to \omega}\lim _{k\to \infty}\left ( q_n(C_{2k}^{k})^2 \frac {k\sin (\pi \frac {p_n}{q_n})}{\sinh \left(k \sin (\pi \frac {p_n}{q_n})\right )}\right )^{\frac {1}{2k}},
\end{eqnarray}
from which, the first approximation of the edge of the butterfly spectrum is given by
\begin{eqnarray}
\label{edgeformula}
E_{max}^{(1)}(\omega )=4e^{-\frac {1}{2}\sin (\pi \omega )}\;\;,
\end{eqnarray}
where $0\leq \omega \leq 1$ and the flux is $\Phi =2\pi \omega .$
\begin{figure}[ht]
\label{spectru}
\centering
\epsfig{figure=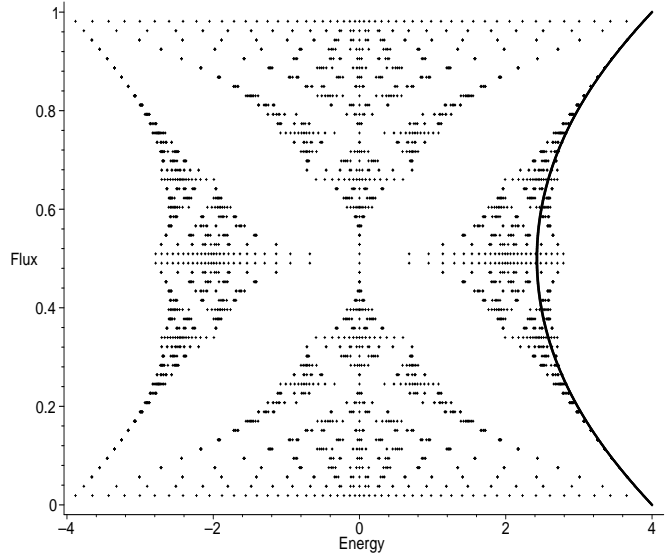,height=10cm, width=8cm,angle=270}
\caption{Spectrum for $q=53$ and the function $4e^{-\frac {1}{2} \sin (\pi \omega )}$, the first approximation for the edge}
\end{figure}

 Figure $2$ presents, for each $p/q,\;q=53\;,p=1,2\cdots 52$, the spectrum $E_j(p/q),\;j=1\cdots q$ on a horizontal line. The vertical coordinate is called Flux ( though we represented only $p/q$ and not $2\pi p/q$). The horizontal coordinate is obviously called Energy. Superimposed on the above described spectrum is the graph of $4e^{-\frac {1}{2} \sin (\pi \omega )}$ as a function of $\omega $, which is the first approximation to the edge of the spectrum. Second order approximations can be obtained by summing the next largest power of $k$, namely all terms that contain $k^{2m-1}.$ It is quite clear that a full understanding of the sequence of polynomials $P_m(k)$ will enable us to decipher the fractal character of the edge of the spectrum.

 At this point it is worthy to mention works connected to the eigenvalues' moments formula, (\ref{Formula}). The geometric interpretation of the trace of the Hamiltonian can be found, for example, in the paper \cite{Bellisard Camacho}. For this, consider the square lattice $Z^2$, where $Z$ are the integer numbers. Let $\Gamma _{2k}$ be the set of all paths of lengths $2k$ which start and end at the origin
$(0,0).$ For such a path $\Gamma $, let $area(\Gamma )$ be the oriented area enclosed by $\Gamma ,$ see \cite{Mingo and Nica} for details. Then we have:
\beq
\label{area}
Tr(H^{2k})=q\sum _{\Gamma \in \Gamma _{2k}}e^{i\pi \frac {p}{q} area(\Gamma )}\;\;.
\eeq
 Compare this expression with the formulas for the eigenvalue moments. For example, for $k=2$ we obtain that $4\cdot 7=28$ represents the number of  paths of lenght $4$ starting and ending at the origin, which enclose zero area. With the interpretation (\ref{area}) for the eigenvalues' moments, it is easy to explain the maximum value $[k^2/4]$ reached by $j$ in (\ref{Mometformula}).
The maximum value for the $area(\Gamma ),$ when $\Gamma $ has a fixed perimeter of $2k,$ is achieved when $\Gamma $ is a square with the edge of length $k/2$ (when $k$ is even), so the area is $k^2/4.$ By the same token, for $k$ odd,  the maximum area is $[k^2/4].$
   
   As $k\to \infty $  the random walks with $2k$ steps, aproaches (using a suitable renormalization, see \cite{Bilingsli}), the 2-dimensional Brownian motion. The probability distribution of the areas enclosed by a planar Brownian motion was computed first by P. L\'{e}vy, \cite{Levy}. L\'{e}vy 's result is 
\beq
\label{Levy}
  E[\exp(ig\;area)]=\frac {g\;area/2}{\sinh (g\;area/2)}\;\;,
\eeq
where $E[\;]$ is the expectation value and $g$ is a real parameter.
Also, in a recent paper \cite{Mingo and Nica}, Mingo and Nica studied the power sums of the areas. What the authors found is that:
\beq
\label{Nica}
\left (\;^{2k}_{\;k}\right)^{-1}\sum _{\Gamma \in \Gamma _{2k}}(area(\Gamma ))^{2m}=R_{2m}(k),\;\;if \;\;k>2m\;\;,
\eeq
where $R_{2m}(k)$ is a rational function in $k.$  The degree of $R_{2m}(k)$ (i.e., the difference of the degrees of the numerator and the denominator of $R_{2m}(k)$ ) is equal to $2m,$ and the leading coefficient of $R_{2m}(k)$, call it $\nu _{2m},$ is generated by:
\beq
\label{sinus}
\frac {z}{\sin z}=\sum _{j=0}^{\infty }\frac {\nu _{2m}}{(2m)!}(2z)^{2m}\;\;.
\eeq
 Both these results are connected with the generating function for the leading coefficients of the polynomials $P_{m}(k),$ (\ref{Generated}). The formula for the eigenvalue moments, (\ref{Formula}) go beyond the L\'{e}vy's formula. From this perspective, a full understanding of the polynomials $P_{m}(k)$ will put in a new light the planar Brownian motion.

\section {Asymptotic Representation of the Product of Bandwidths  for  a Weak Magnetic Field and Szeg\" {o}'s Theorem}

We now aim to compute the asymptotic representation of the product of the bandwidths, $\prod _{n=1}^q\Delta_n,$ for a weak magnetic flux, $\Phi =2\pi \frac {1}{q},$ when $q\to \infty .$ We will use the inequality (\ref{ProdBand}) together with the asymptotic representation of the Hankel determinant, (\ref{Hankel}). The entries of the Hankel determinant are the moments $s_{2k}$, see (\ref{sk}).   
Up to the first order of approximation, (\ref{first}), the moments are: 
\beq
\label{s2k}
s_{2k}=q\left (\;^{2k}_{\;k}\right)^2 \frac {k\sin(\frac {\pi }{q})}{\sinh (k\sin(\frac {\pi }{q}))}\;\;.
\eeq
To go further, and because of lack of knowledge of the polynomials $P_m(k)$, I assume that ( for the case of a week flux), the second and all other orders of approximation are much smaller than the first order. Morover, because
\beq
\label{ineq0}
C_1<\frac {k\sin(\frac {\pi }{q})}{\sinh (k\sin(\frac {\pi }{q}))} e^{\frac {k}{2q}}<C_2,\;\;\; k=1,2,\cdots q-1\;\;,
\eeq
I will consider that the influence of the first approximation term, i.e.
\beq
\label{approx}
 \frac {k\sin(\frac {\pi }{q})}{\sinh (k\sin(\frac {\pi }{q}))}\;\;,
\eeq
on the asymptotic formula is of the same order of magnitude as given by the term $e^{-\frac {k}{2q}}.$ In (\ref{ineq0}) $C_1\sim \pi e/\sinh(\pi )=0.739$ and $C_2\sim e^{1/\pi }/\sinh(1)=1.169 .$ Thus the moments $s_{2k}$ to be used to find the assymtotic formula for the Hankel determinant are:
\beq
\label{s2k1}
s_{2k}=q\left (\;^{2k}_{\;k}\right)^2 e^{-\frac {k}{2q}}\;\;,
\eeq

and it follows that
\beq
\label{HankelBinom}
\sigma_{q-1}=q^q e^{-q+1}\left|\begin{array}{ccccccc}
          1     & 0&\left (\;^{2}_{1}\right)^2  &0      & \left (\;^{4}_{2}\right)^2  & \cdots    &        \\
          0&\left (\;^{2}_{1}\right)^2    &0 & \left (\;^{4}_{2}\right)^2   &  \vdots &       &         \\                                                                          
          \left (\;^{2}_{1}\right)^2    &0 & \left (\;^{4}_{2}\right)^2   &\vdots     &   &    &       \\
           0 & \left (\;^{4}_{2}\right)^2   &\vdots   &       &  & &      \\
          \left (\;^{4}_{2}\right)^2   &\vdots  &    &  &   &    &         \\

          \vdots   &      &  &   &    &  &  
           
         \end{array}\right |\;\;.
\eeq

 The dimension of $\sigma _{q-1}$ is $q.$ To find the asymptotic representation of $\sigma _{q-1},$ as $q\to \infty ,$ we shall employ a theorem of  G.Szeg\" {o}. One version of this theorem, \cite {Johansson}, gives the asymptotic behaviour of a determinant 
$D_{n-1}(f):=\det(b_{\mu \nu })_{0\leq \mu ,\nu \leq n-1}$ whose  
 generic elemet  can be expressed as  
\beq
\label{bmunu} 
b_{\mu \nu}=\int _{-1}^1 x ^{\mu+\nu} f(x)dx\;\;,
\eeq
 that is to say, the entries of the determinant can be expressed as the moments of a function $f(x).$
 
 The theorem stays that, as $n\to \infty ,$
\beq
\label{Szego}
D_{n-1}(f)=D_{n-1}(1)\exp \left(nc_0+\frac {1}{8}\sum _{j=1}^{\infty }jc_j^2+O(1)\right )\;\;.
\eeq

where $c_k,\;\;k=0,1\cdots ,$ are the Fourier coefficients of the function $\ln(f(\cos \theta )):$
\beq
\label{Fourier}
c_k=\frac {1}{2\pi }\int _{-\pi }^{\pi }e^{-ik\theta }\ln(f(\cos \theta ))d\theta\;\; .
\eeq

To apply this theorem we need the function $f.$ The generic term in the Hankel determinant $\sigma _{q-1}$ can be expressed as an integral as followes: 
\beq
\label{Elliptic}
b_{\mu \nu }=\frac{2}{\pi ^2}4^{\mu +\nu } \int _{-1}^1 x^{\mu +\nu }K^{'}(x)dx\;\;,
\eeq
where $K^{'}(x)$ is the complete elliptic integral of the first kind:
\beq
\label{CompletEllitic}
K^{'}(x)=\int _0^{\frac {\pi }{2}}(1-x^{'2}\sin \theta )^{-1/2}d\theta =\frac {\pi }{2}\;_2F_1\left(\frac {1}{2},\frac {1}{2};1;x^{'2}\right)\;\;.
\eeq
 Here $x^2+x^{'2}=1$ and $_2F_1(a,b;c;x)$ is the Gauss' hypergeometric series \cite{Abramowitz}.
 
 The function $f$ is then 
\beq
\label{fK}
f(x)=K^{'}(x)\;\;.
\eeq

 The determinat $D_{n-1}(1)$ can be computed exactly and is given by the following product, using  
formula $(2.2.15)$ in \cite{Szbook}:
\beq
\label{AsSz}
D_{n-1}(1)=2\prod _{k=1}^{n-1}\left(k+\frac {1}{2}\right )^{-1}\left (\frac {1}{2^k} C_{2k}^k\right )^{-2}\;\;. 
\eeq
 For large value of $n$
$$D_{n-1}(1)= 2^{-n(n-1)+O(1)}\;\;.$$

Combining with the result of Szeg\" {o}'s theorem, the asymptotic representation of
$\sigma_{q-1}$ reads as:
\beq
\label{Assigma}
\sigma_{q-1}=q^q e^{-q+1}\left (\frac {2}{\pi ^2}\right )^q 4^{q(q-1)}D_{q-1}(1)\exp\left(qc_0+\frac {1}{8}\sum _{j=1}^{\infty }jc_j^2+O(1)\right)\;\;, 
\eeq
 Here $c_0=0.729.$
 
From this we get, as $q\to \infty$, for  a weak magnetic field $(\Phi =2\pi \frac {1}{q}),$
\beq
\label{Prodas}
\prod_{n=1}^{q}\Delta _n \sim 2^{-q^2}q^{-q} \exp(-qd_0+d_1+O(1))\;\;,
\eeq
where $d_0,\;d_1$ are constant which cannot be made precise because of the innequality we started with, (\ref{ProdBand}), and because of the approximations we used, (\ref{s2k}) and (\ref{s2k1}). 

Suppose now that the bandwidths are written as an exponential:
\beq
\label{expdelta}
\Delta _{n}=e^{-\mu _n q}\;\;.
\eeq
From the above definition we obtain the asymptotic formula for the average value  $\mu _n:$
\beq
\label{summu}
\frac {\sum _{n=1}^q \mu _n}{q}=\ln 2+\frac {\ln q}{q}-\frac {d_0}{q}+O\left(\frac {1}{q}\right)\;\;.
\eeq
 At the limit:
\beq
\label{log2}
<\mu >:=\lim _{q\to \infty }\frac {\sum _{n=1}^q \mu _n}{q}=\ln 2\;\;.
\eeq

\section {Conclusions}

 This paper contains asymptotic results on the bandswidths in the Hofstadter spectrum for 2-dimensional electrons in a magnetic field. Besides definite results, it also leaves some open questions.
 The important results are: asymptotic formula (\ref{Prodas}), the formula for the edge of the spectrum (\ref{edgeformula}), and the conjecture on the general formula for the eigenvalue moments (\ref{Formula}).

The open questions are: What is the general formula for the polynomials $P_{m}(k)?$ How to obtain the fractal structure of the edge of the spectrum from the eigenvalue moments? How to find the asymptotic representation for the product of the bandwidths for every flux, using Szeg\" {o} formula?

\section{Acknowledgments}
I am grateful to P.B Wiegmann for introducing me to the subject and for his many valuable comments and ideas. I would like to express my gratitude to B. Simon for helpful discussions and for the fruitful scientific environtment he created during the time this paper was written. Special thanks I owe to A.G. Abanov, P. Di Francesco, Y. Last, V. Sahakian, A. Soshnikov and J. Talstra, for stimulating discussions.
\section{Appendix}
I list here the first eight polinomials $P_m(k)$, which appeared in eigenvalue moment formula (\ref{Formula}). The following formulas are conjectured from the first 25 eigenvalue moments (\ref{eigenmoments}). 

\begin{eqnarray}
 P_1\!\!\!\!&=&\!\!\!\!-3^{-1}\\ \nonumber \\
 P_2\!\!\!\!&=&\!\!\!\!90^{-1}(7k^2-4k-15)\\ \nonumber \\
 P_3\!\!\!\!&=&\!\!\!\!-1890^{-1}(31k^4-60k^3-101k^2+66k+280)\\ \nonumber \\
 P_4\!\!\!\!&=&\!\!\!\!37800^{-1}(127k^7-1044k^6+2246k^5+328k^4+
7k^3-\\ \nonumber &\quad& \quad \quad 12244k^2-220k+25200)\\ \nonumber \\
P_5\!\!\!\!&=&\!\!\!\!-3742200^{-1}(2555k^9-31887k^8+137946k^7-243774k^6+\\ \nonumber 
&\quad& \quad \quad  290499k^5-796527k^4+647416k^3+1661436k^2-\\ \nonumber \quad &\quad& \quad \quad 95184k-3991680)\\ \nonumber \\
P_6\!\!\!\!&=&\!\!\!\!10216206000^{-1}(1414477k^{12}-32155043k^{11}+300761927k^{10}-
\\ \nonumber
&\quad &\quad \quad 1517115007k^9+4720753473k^8-10693488621k^7+20222071853k^6-\\ \nonumber
&\quad &\quad \quad25923159133k^5+11996284390k^4+
11841173876k^3+ \\ \nonumber
&\quad &\quad \quad45697115160k^2+12156530400k-90810720000)\\ \nonumber \\P_7\!\!\!\!&=&\!\!\!\!-10216206000^{-1}(286685k^{14}-8595813k^{13}+110951109k^{12}-\\ \nonumber
&\quad &\quad \quad 819736201k^{11}+3945073115k^{10}-13604875839k^9+ \\ \nonumber
 &\quad &\quad \quad 36023850727k^8-73323185763k^7+ 108429092220k^6- \\ \nonumber
&\quad &\quad \quad 120643403168k^5+122711359264k^4-65709283536k^3-\\ \nonumber &\quad &\quad \quad 63479070720k^2-19852750080k+186810624000)\\ \nonumber \\
P_8\!\!\!\!&=&\!\!\!\!20841060240000^{-1}
(118518239 k^{17}-5242493627k^{16}+104055035258k^{15}-  \\ \nonumber 
 &\quad &\quad \quad 1233596237660k^{14}+
9844400576738k^{13}-56735776354034k^{12}+ \\ \nonumber
&\quad &\quad \quad 247874192827336k^{11}-847184326535620k^{10}+2294971824780007k^9-\\ \nonumber
&\quad &\quad \quad 4924658969019451k^8+8359347851375254k^7-11207926976651080k^6+ \\ \nonumber
&\quad &\quad \quad 11409224332806816k^5-7994766402003888k^4+4489556615965152k^3-\\ \nonumber
&\quad &\quad \quad 3442931101232640k^2-806985556876800k+4668397493760000)
\end{eqnarray}

\end{document}